\documentclass[letterpaper]{article} 
\usepackage{aaai24}  
\usepackage{times}  
\usepackage{helvet}  
\usepackage{courier}  
\usepackage[hyphens]{url}  
\usepackage{graphicx} 
\usepackage{natbib}  
\usepackage{caption} 
\usepackage{algorithm}
\usepackage{algorithmic}

\usepackage{newfloat}
\usepackage{listings}
\DeclareCaptionStyle{ruled}{labelfont=normalfont,labelsep=colon,strut=off} 
\floatstyle{ruled}
\newfloat{listing}{tb}{lst}{}
\floatname{listing}{Listing}
\usepackage{xcolor} 

\title{Distributed Monitoring for Data Distribution Shifts in Edge-ML Fraud Detection}

\author {
    Nader Karayanni\textsuperscript{\rm 1}, Robert J. Shahla\textsuperscript{\rm 2}, Chieh-Lien Hsiao\textsuperscript{\rm 1}
}
\affiliations {
    \textsuperscript{\rm 1}Columbia University\\
    \textsuperscript{\rm 2}Technion\\
    nader.karayanni@columbia.edu, shahlarobert@campus.technion.ac.il, ch3689@columbia.edu
}

\usepackage{bibentry}

\begin{document}

\maketitle

\begin{abstract}

The digital era has seen a marked increase in financial fraud. edge ML emerged as a promising solution for smartphone payment services fraud detection, enabling the deployment of ML models directly on edge devices. This approach enables a more personalized real-time fraud detection.
However, a significant gap in current research is the lack of a robust system for monitoring data distribution shifts in these distributed edge ML applications. Our work bridges this gap by introducing a novel open-source framework designed for continuous monitoring of data distribution shifts on a network of edge devices.
Our system includes an innovative calculation of the Kolmogorov-Smirnov (KS) test over a distributed network of edge devices, enabling efficient and accurate monitoring of users behavior shifts.
We comprehensively evaluate the proposed framework employing both real-world and synthetic financial transaction datasets and demonstrate the framework's effectiveness.

\end{abstract}
\section{Introduction}
In recent years, smartphone payment services such as Apple Pay, smart wallets, and banking apps have revolutionized financial transactions, offering unprecedented convenience to users. However, these advancements have significantly increased fraud risk, by methods including identity theft, phishing, and smartphone theft \cite{Mobile-Wallet-Threats, Mobile_payment_security_threats_and_challenges}.

To combat these risks, payment service providers (referred to as operators) increasingly rely on Machine Learning (ML) to identify and prevent fraudulent activities with high accuracy \cite{state-of-the-art-review-of-machine-learning-techniques-for-fraud-detection}. Recently, there has been increased adoption of on-device ML, also known as edge ML, which is a paradigm where lightweight ML models are deployed directly on edge devices such as smartphones, sensors, and IoT devices. One advantage to this approach is the lower latency in detecting fraud since the inference doesn't require traveling over the network. Another advantage is enabling the deployment of a personalized fine-tuned ML model per user, significantly enhancing the understanding of user-specific spending patterns and other relevant variables, therefore opening new avenues in improving the detection of fraudulent activities~\cite{Anomaly-Detection-Techniques:ON-DEVICE-ML-FOR-FRAUD-DETECTION, Fraud-Detection-Based-on-Individual-Behavior}. 

ML-based fraud detection systems often operate under a critical assumption: the data used for training these models closely mirrors the data encountered in real-world applications. However, this assumption doesn't always hold. Data distribution shift is a phenomenon where the statistical properties of the data distribution change over time. This change can possibly lead to a significant undermining of the  ML model's performance, as the encountered data no longer reflects the distribution on which the ML model was originally trained~\cite{LearninginthePresence-of-Drift}, denoted as \textit{originial training data}. In the context of financial transactions, as an example, consumer behavior changes lead to shifts in spending patterns that ML models may not be trained to handle, invalidating the assumption that the ML model original training data mirrors the real-world encountered data~\cite{Credit-card-fraud-detection-and-concept-drift-adaptation-with-delayed-supervised-information}.

Identifying the distribution shift is typically done via a proactive monitoring process, where the newly incoming data distribution is continuously monitored and compared against the original training data distribution. This monitoring empowers the operators to promptly identify instances of data distribution shift and take action accordingly, preventing prolonged performance degradation of the ML models.

In the context of edge ML,  monitoring shifts in data distribution becomes more challenging. The newly incoming data is distributed across a network of edge devices, and new samples are continuously being generated at each device. Identifying distribution shifts across the edge devices requires consideration of the data generated by the distributed edge devices.
A monitoring system deployed in the operator's cloud cluster is tasked with aggregating and monitoring the data distribution from these devices. The na\"ive solution is for the edge devices to send each new sample to the cloud. However, this approach is impractical due to its overhead for both the clients and the operator. The transmission of a vast number of samples would demand excessive bandwidth and runtime at the edge devices. Furthermore, the operator would need substantial processing power to handle the volume of incoming data.

The challenge mentioned above remains underexplored. This work addresses this gap, We introduce a novel framework for continuous monitoring of data distribution shifts in such environments. Our primary contribution lies in the design and implementation of this framework, with a specific emphasis on its application in edge ML fraud detection scenarios. We conduct thorough evaluations of our framework highlighting its efficiency and effectiveness.

\section{Background and Related Work}

\subsection{Edge ML for Fraud Detection}
The detection of fraudulent transactions is a crucial task in the finance industry. Fraud Detection Systems (FDS) heavily rely on ML as a backbone for classifying fraud. Researchers and industry leaders alike employ ML-based methods, with neural networks being the most adopted method~\cite{state-of-the-art-review-of-machine-learning-techniques-for-fraud-detection}. \citet{Credit-card-fraud-detection-using-artificial-neural-network} have demonstrated the effectiveness of neural networks in classifying transactions as fraudulent or legitimate. Similarly, FDS industry leaders such as~\citet{featurespace} and~\citet{complyadvantage}, heavily rely on ML for fraud detection.

Edge ML refers to the deployment of ML models on edge devices, such as smartphones, IoT devices, and other computing resources, rather than in the cloud. This approach provides many benefits, such as enabling the use of a personalized fine-tuned ML model per edge device, minimizing latency, conserving bandwidth, and preserving privacy. However, the primary drawback of edge ML lies in the limited resources of edge devices and the absence of centralized performance monitoring for the algorithms \cite{A-survey-on-deploying-mobile-deep-learning-applications}.

Edge ML is attractive in the context of mobile payments fraud detection. \citet{Personalized-Detecting-Fraud} have shown an effective way to build a personalized ML model based on a questionnaire to assess the consumer's behavior. Similarly \citet{Fraud-Detection-Based-on-Individual-Behavior} provided a fraud detection algorithm that leverages individual behaviors to improve accuracy. Beyond personalization, edge ML provides other advantages for financial applications. For example, \citet{FFD} have utilized the privacy provided by edge ML to develop a federated learning-based FDS that enables banks to collaboratively learn and improve fraud detection models while keeping their data localized and secure.

\subsection{Handling Data Distribution Shift}
Data distribution shift refers to the phenomenon where the distribution of the newly observed data differs from the original training data of the ML model. This shift can lead to a decline in the model's performance as the original training data no longer reflects the current real-world data distribution~\cite{LearninginthePresence-of-Drift}. In the dynamic world of financial transactions, and with the continuous changes in the users and market behaviors, the phenomenon of data distribution shift poses a significant challenge ~\cite{Realistic-Modeling-and-a-Novel-Learning-Strategy}. 

A key step to detect data distribution shift is the continuous monitoring of incoming data. This process entails comparing the distribution of the new data against the distribution of the original training data. A divergence between these two distributions signals the occurrence of data distribution shift. Detecting such shift is vital as it helps the operator recognize potential declines in model performance, and take timely and informed actions accordingly.

For cloud-based ML solutions, monitoring the incoming data for identifying data distribution shifts is an essential step \cite{AWS:DRIFT_DETECTION, GCP:DRIFT_DETECTION}. \citet{nannyml} is a framework dedicated to monitoring cloud-based ML models' post-deployment. Monitoring data distribution shift is one of the critical methods NannyML relies on to help operators. Besides monitoring the incoming data, assessing data shifts can be done by utilizing the ML model's outputs. For example,~\citet{Confidence-for-drift-detection} and~\citet{nazar} utilize the outputted probabilities of the neural network, where systematic low model confidence is correlated with distribution shift.


\begin{figure}[t]
    \centering
    \includegraphics[width=0.9\columnwidth]{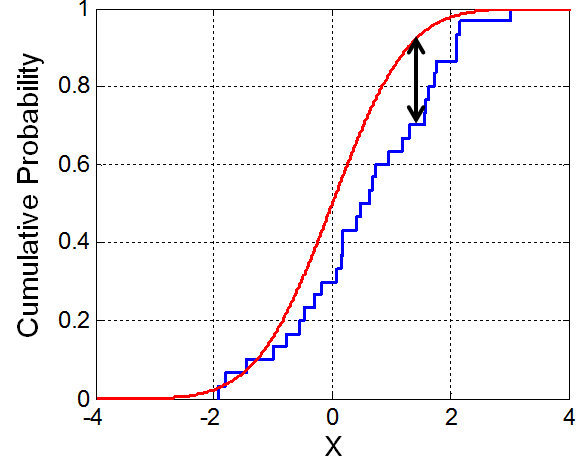} 
    \caption{KS-statistic visualization \cite{enwiki:ks}.}
    \label{KS-statistic}
\end{figure}

\subsection{Two Distributions Comparison}
A widely adopted test for comparing two distributions is the Goodness of Fit (GoF) test. A GoF test can be used to statistically determine how well the distribution of new data aligns with the distribution of the original training data. This involves quantifying the differences between the two distributions. Among the various GoF tests, the Kolmogorov-Smirnov (KS) test stands out as a popular GoF test in ML applications ~\cite{KS-Original-Paper, KS2016fast}. This non-parametric test is favored for its distribution-free nature. The KS test operates by calculating the maximum distance between the Cumulative Distribution Function (CDF) of two distributions (Figure \ref{KS-statistic}). Formally, if we denote by $F_1$ and $F_2$ the CDFs of two distributions, then the KS statistic is given by:  \[ KS(F_1,F_2) = \sup_x |F_1(x) - F_2(x)| \]
A KS statistic close to 0 suggests that the two compared distributions are aligned. Conversely, as the KS statistic increases, so does the indicated divergence between the distributions, where 1 is the maximum KS statistic.

Recent work has been expanding the original KS test to suit the current ML solutions. \citet{streaming-KS} and~\citet{Online-Drift-Detection-Using-Incremental-KS} provide algorithms based on the KS test, which are designed for monitoring streaming data. 
In~\citet{big_data_ks}, an extended version of the KS test tailored for scenarios involving big data was introduced. While the paper briefly touched upon the potential extension to a distributed setting, the details were not discussed.

In this work, we focus on utilizing the KS test as our GoF test for monitoring data distribution shifts. this decision is due to the high adoption and proven effectiveness of the KS test. As an essential part of our framework, we propose a novel calculation of the KS test in a distributed setting and evaluate its efficiency and accuracy.

\subsection{$t$-digest Data Structure}
The $t$-digest~\cite{t-digest} data structure is designed to dynamically represent a distribution based on continuously generated samples. It represents the distribution with a series of centroids, each summarizing a subset of the data. The $t$-digest creates and updates the centroids dynamically according to the new samples' distribution, Providing finer resolution in areas with a high density of data points and coarser resolution in areas of lower density. This adaptability makes it robust in representing different data distributions with minimal memory footprint.

The $t$-digest offers two knobs, the $delta$ knob controls compression, affecting both the size and accuracy of the data structure. A lower $delta$ results in higher compression but lower accuracy. The $k$ knob limits the size of centroids, balancing quantile precision with memory usage.

\section{Proposed Framework}
In this section, we introduce our framework designed to assist in effectively monitoring data distribution shifts in edge ML deployments. Our framework provides continuous distributed edge data aggregation and calculation of the KS test for the aggregated data. A significant challenge in the framework is calculating the KS statistic across a distributed network of devices, without transmitting the entire data to the cloud. The solution is designed to accurately and efficiently calculate the KS statistic while minimizing storage and network bandwidth usage on edge devices.

To overcome the above challenges, each edge device maintains a compact data structure, which encapsulates the distribution of samples generated within a specific time window. Its primary function is to represent the CDF of the data observed by the edge device using sketches optimized for minimal storage overhead, efficiency, accuracy, and the ability to be merged with little to no loss in accuracy.

To meet these needs, the $t$-digest data structure is used. It is efficient both in its memory footprint and computation overhead, accurate,  and can be easily merged across devices to yield a global view of the data distribution. Thereby facilitating the distributed calculation of the CDF. The merge is carried out by iteratively combining the centroids of each $t$-digest, and adjusting their weights and positions to reflect the combined data distribution. The merged $t$-digest accurately represents the combined distribution represented by the two original $t$-digest objects.

\subsection{High-level Overview}
\begin{figure}[h]
\centering
\includegraphics[width=1\columnwidth]{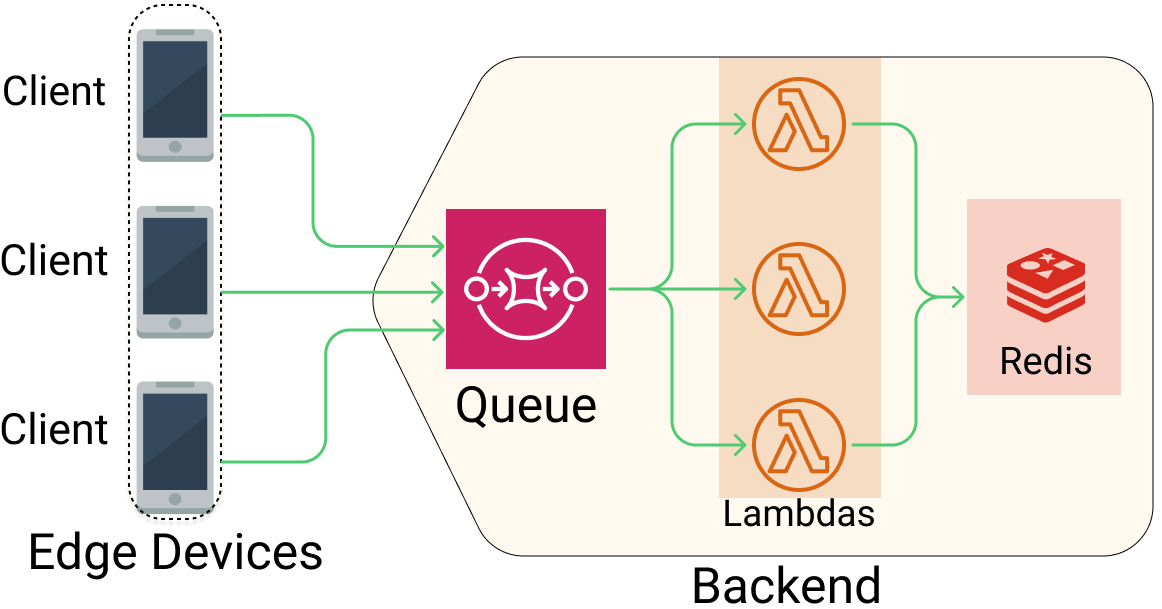}
\caption{Framework overview}
\label{system overview}
\end{figure}

Each edge device stores a $t$-digest object, which is updated with every new sample generated by the device. When $d$ samples are generated, the device sends the $t$-digest object to the operator. Once the objects from the client devices are received, the operator aggregates them into a unified $t$-digest object. This unified object reflects the data distribution generated by the entire network of devices. Consequently, the KS statistic can be calculated once a sufficient number of samples have been received by the operator. The system architecture includes two primary components (Figure~\ref{system overview}).

\begin{itemize}
    \item \textbf{Client:} A client is deployed on each edge device and continuously tracks newly generated samples. The goal of a client is to encapsulate the seen samples' distribution efficiently, and eventually, send this data to the backend.
    \vspace{2mm}
    \item \textbf{Backend:} The backend manages a queue to which the clients' data is added. In addition, it analyzes the collected data and implements the monitoring.
\end{itemize}

\section{Framework Implementation}
The framework is developed in Python and is available as an open-source project on GitHub\footnote{\url{https://github.com/karayanni/Distributed-KS-Test}}. The repository includes all necessary code, detailed setups, and instructions for replicating our experiments.

\subsection{Client}
The client component of the framework is the first layer of the distributed data monitoring system. This Python-based client code is lightweight, and designed to be deployed on edge devices. The implementation of $t$-digest in~\citet{t-digest-python} was used. At its core, the client code is responsible for managing the $t$-digest object which represents the sampled CDF by the client. For every sample generated by the device, the $t$-digest object data is updated. Note that updating the $t$-digest can also be done in batches by grouping new samples and updating the $t$-digest object once per batch.

A key configurable parameter is the number of samples inserted into the $t$-digest object, before transmitting it to the operator, denoted as $d$. Once the number of processed samples surpasses this threshold, the client initiates transmitting the data structure, along with its associated metadata to the operator. This metadata contains information such as the used ML model identifier, user payment location, age, etc.

\subsection{Backend} 
The backend architecture is designed in a serverless fashion, leveraging AWS services for scalability and resource efficiency. The backend consists of the following components.
\begin{description}
\item[Queue]
Simple Queue Service (SQS) from \citet{SQS}, used as the data queue.
All data received from the clients is inserted into this queue in a FIFO manner.

\item[Redis]
To compute the aggregated CDF from multiple clients, a $t$-digest object is maintained in a \citet{Redis} instance, chosen for its fast in-memory data handling capabilities. Multiple CDFs can be monitored by maintaining different $t$-digest objects accordingly.

\item[Lambda]
Upon receiving data from a client, SQS triggers a Lambda function dedicated to processing this data. The purpose of this function is to:
\begin{enumerate}
    \item Identify the proper $t$-digest object(s) (if there are multiple), based on the provided metadata.
    \item Retrieve the relevant $t$-digest object(s).
    \item Merge the newly received $t$-digest object with the relevant existing one(s). optimistic locking is used for concurrency control.
\end{enumerate}
\end{description}

\begin{figure}[h]
\centering
\includegraphics[width=1\columnwidth]{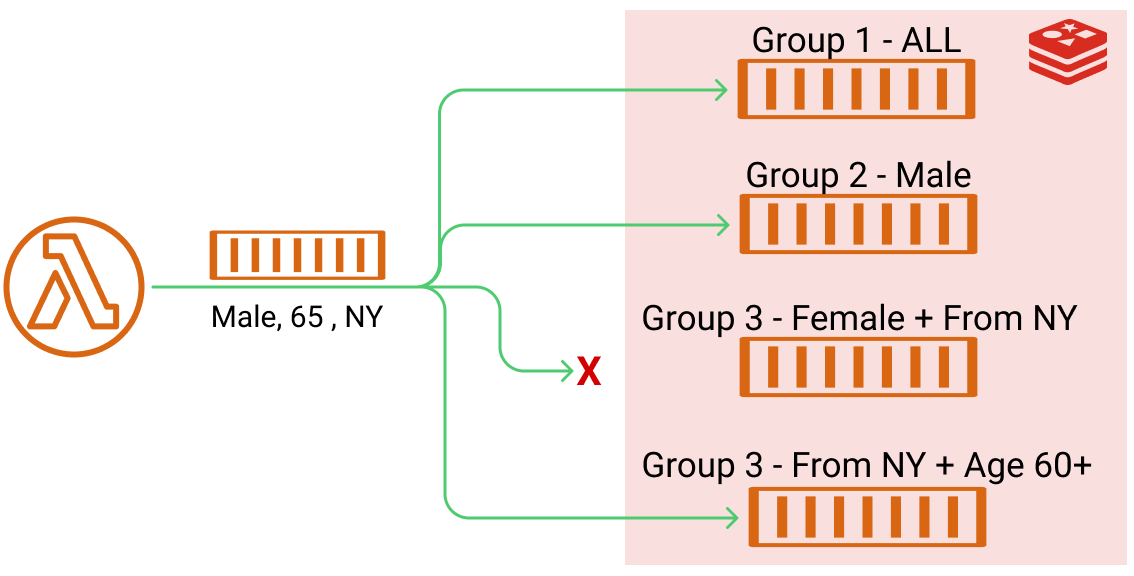} 
\caption{Handling queue message.}
\label{handle queue message server}
\end{figure}

Figure~\ref{handle queue message server} presents an example of handling a queue data insertion, where 4 different distributions are monitored, for 4 population types. An object corresponding to a male, 65 years old from New York (NY) is received. Only the objects of groups 1, 2, and 4 are updated in this case.

After merging the newly received $t$-digest object, the Lambda function assesses if the updated $t$-digest object(s) represents a sufficient number of samples, surpassing a pre-defined threshold, denoted as \textit{aggregation size}. If so, for each such object, it performs the KS test, comparing the current batch distribution with the original training dataset distribution. The operators can then integrate alerting logic for monitoring a distribution shift based on a KS score threshold. Following the KS test, the $t$-digest object is reset and is ready to repeat the aggregation process.

\section{Evaluation}
In this section, we rigorously evaluate our framework, addressing the following key aspects:
\begin{enumerate}
    \item The accuracy of our distributed KS statistic calculation.
    \item The effect of our KS calculation on false positive and false negative alerting of distribution shift.
    \item The runtime overhead of the server.
    \item The runtime overhead for each client.
\end{enumerate}

\subsection{Datasets}
Our framework is designed to assist operators in detecting shifts in their users' financial data distribution. To accommodate the expected use cases, we conducted experiments on datasets containing financial transactions.
We use a real-world dataset containing 284,807 credit card transactions from 2013, as described in \cite{real-world-ds}, denoted as \textit{real-world dataset}. We also use two synthetic data sets, collectively comprising over 8 million simulated financial transactions \cite{generatedfraud212, generatedfraud186}, denoted as \textit{synthetic datasets}. These datasets provide detailed insights into transaction amounts, cardholder information, and fraud indicators. Closely mirroring the data expected to be encountered by the operators.

\subsection{Accuracy}
In real-world scenarios, aggregating all the data from edge devices to compute the KS statistic is often impractical. This is why it's crucial for our framework to provide a precise estimation of the KS statistic, reflecting the collective data observed by the edge devices. In this part of the evaluation, we thoroughly compare the accuracy of the proposed distributed method, denoted as \textit{T-Digest-KS}, against the scenario where complete data is accessible, denoted as \textit{Optimal-KS}. The tests cover the impact our distributed computation has on false positive and false negative identifications of data shifts.
In all our experiments, we use the default $t$-digest configurations, specifically setting $delta$ to 0.01 and $K$ to 25.

\subsubsection{Real-World Dataset Evaluation}
This experiment evaluates the T-Digest-KS using the real-world dataset. The objective is to determine how accurate the T-Digest-KS calculation of the KS statistic is, with different data shift degrees.

Due to the dataset's small size, a random 100,000 transactions are used as the original data. Another 100,000 transactions, selected randomly from the remaining pool, served as our newly encountered data, representing newly encountered transaction patterns. To simulate shift, we introduced random increases in the spending patterns with varying degrees of increase to result in different levels of shift. For each degree of shift, the process was repeated 15 times.

\begin{figure}[h]
\centering
\includegraphics[width=0.9\columnwidth]{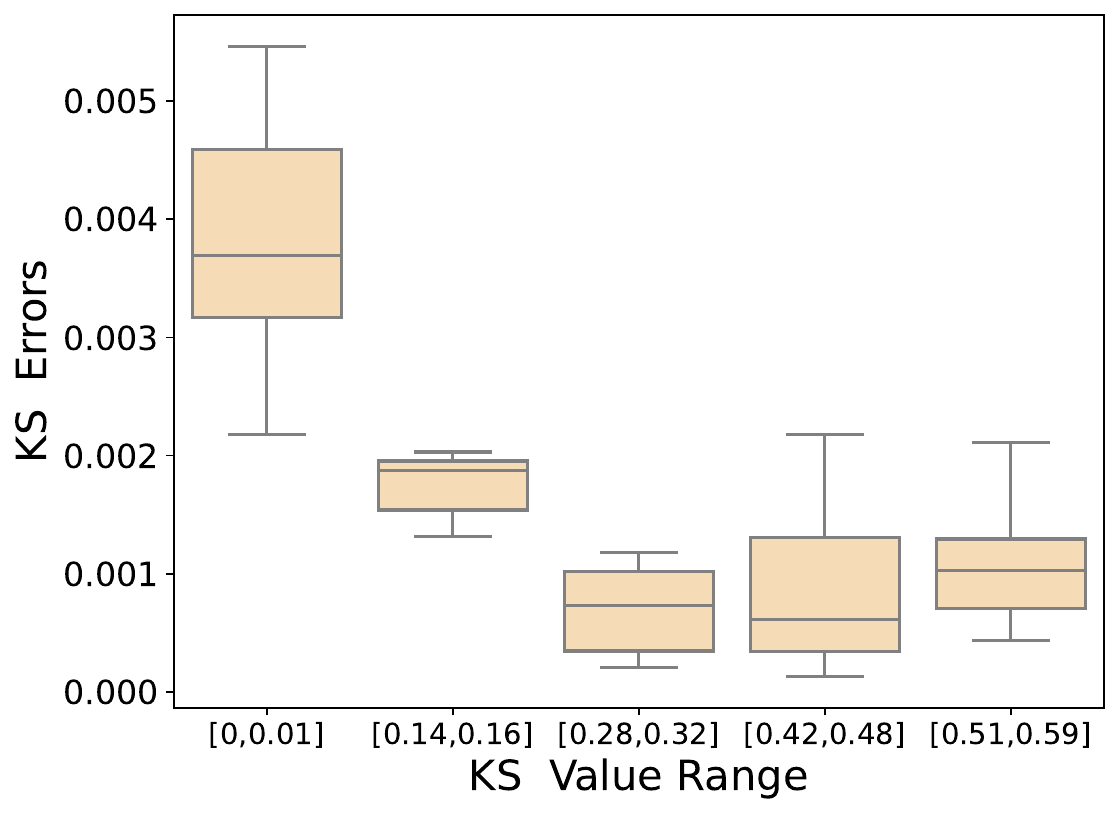}
\caption{Real-world dataset accuracy with different shifts.}
\label{fig:real-world-test}
\end{figure}

Figure~\ref{fig:real-world-test} depicts the experiment results. The Y-axis represents the absolute difference between the KS statistic calculated by T-Digest-KS and Optimal-KS. The X-axis represents the KS value range of the Optimal-KS across the repeated runs.
The results indicate a consistent low error in the T-Digest-KS estimations. The median error remained below 0.004. Interestingly, in cases with slight to significant shifts, T-Digest-KS demonstrated even higher accuracy, with a median error rate below 0.002. The explanation for this phenomenon can be attributed to the lossy-ness of the $t$-digest data structure, and the definitions of the KS statistic, which identifies the greatest divergence in the CDFs of two distributions.

In instances with no shift, the difference between the two CDFs is predominantly zero, and the $t$-digest's lossy-ness becomes the primary factor in determining the discrepancy between the CDFs. However, in scenarios with data shifts, the differences in CDFs are more pronounced. $t$-digest's lossy-ness is comparatively less impacting because it affects a smaller proportion of critical comparisons (points where the difference in the CDFs is a potential candidate to be the KS statistic).

\subsubsection{Immediate Shift in Users Behavior}
An immediate shift is a distribution shift due to a certain percentage of users drastically altering their behavior. Such scenarios are particularly relevant for fraud detection ML models. For instance, a model developed using data prior to COVID-19 might not perform accurately during the pandemic where immediate distribution shifts occur to consumer behavior \cite{Consumer-Behaviour-during-Crises}.
We assess the accuracy of our system using the synthetic datasets mentioned above, which emulate distinct yet realistic spending scenarios. We harness the large dataset size to calculate the KS statistic using 500,000 samples per test. The test involves five varied percentages of users exhibiting immediate shifts, repeating each test 15 times.

\begin{figure}[h]
\centering
\includegraphics[width=1\columnwidth]{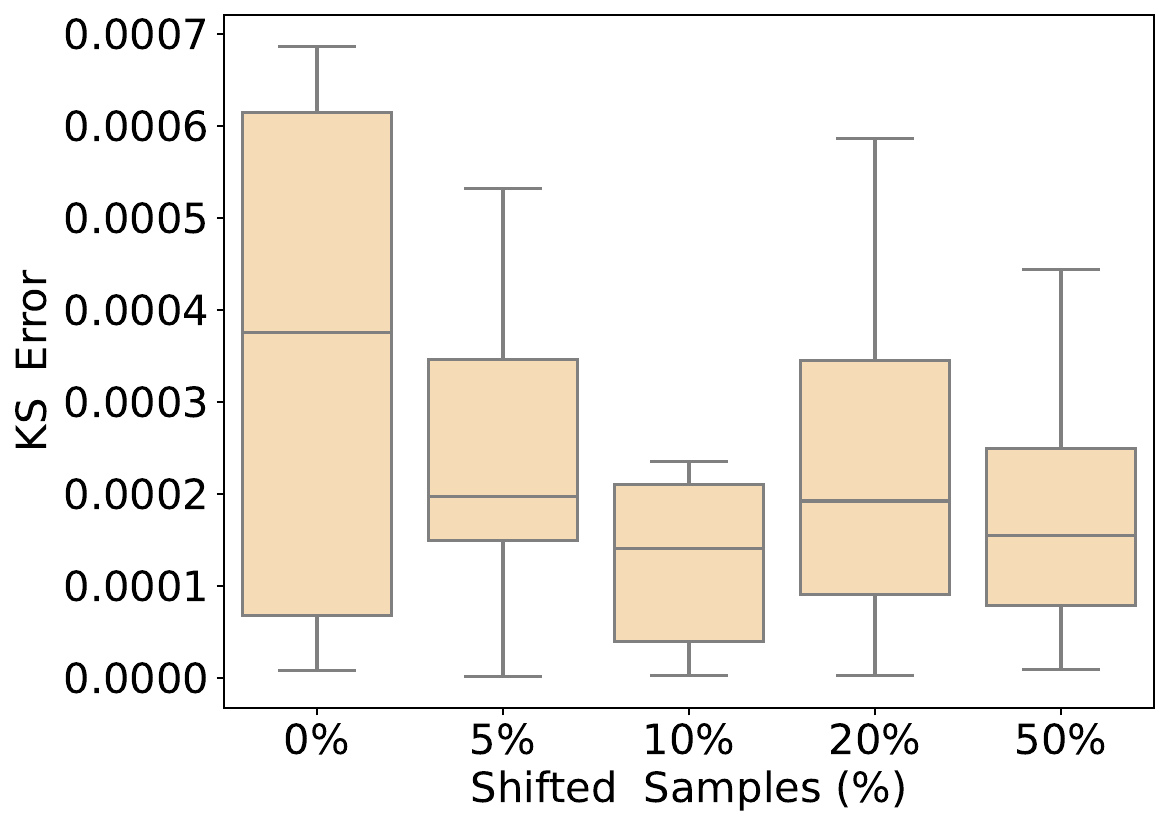}
\caption{Accuracy with a percentage of shifted users.}
\label{fig:sudden_drift}
\end{figure}

As depicted in Figure \ref{fig:sudden_drift}, T-Digest-KS shows remarkable accuracy in this set of tests. The enhanced performance can be attributed to three key differences: the increased number of samples per test, differences in the distribution characteristics, and the nature of the data distribution shift.

\subsubsection{Takeaways}
Building on the findings from the two experiments, T-Digest-KS exhibits a notably low error in estimating the KS statistic. In standard practice, a KS statistic exceeding the range of 0.03-0.05 is often indicative of a significant change in data distribution, as established in \cite{KS-Original-Paper}.

Our comprehensive evaluation underscores that when the KS statistic remains below the 0.01 threshold, the median error in T-Digest-KS estimations is consistently under 0.004 across both tests and the max error measured is below 0.006. This low error significantly minimizes the likelihood of false positives in identifying distribution changes. Moreover, in instances where the KS statistic surpasses the 0.01 threshold, T-Digest-KS demonstrates an even lower error margin. This heightened accuracy further reduces the risk of false negatives.

These results highlight the reliability and precision of T-Digest-KS in detecting shifts in data distribution, ensuring a high degree of confidence in its practical application, particularly in scenarios where accurate detection of data shifts is critical.
 
\subsection{Runtime}
This section assesses the runtime associated with adopting our framework of maintaining a $t$-digest object on each edge device and merging them in the cloud, denoted as \textit{T-Digest-Merge}. As a baseline, a batch-streaming approach was employed. In this method, clients transmit data to the server in batches, without maintaining a $t$-digest object, and the samples are then inserted to the $t$-digest object in the cloud. This method is denoted as \textit{T-Digest-Stream}.

To ensure a fair comparison, the batch size in T-Digest-Stream was set to 1,000 samples, mirroring our system's memory footprint on the client side. This consideration is essential for maintaining a balance between performance and resource consumption on client devices.

We examine the overhead for both the backend server and the client code running on the edge devices. All experiments use random samples from the real-world dataset. The results were averaged across 10 runs. The tests were performed on a system with an 11$^\mathrm{th}$ Gen Intel Core i7 processor, 3.00GHz, and 16GB of memory. To simulate cloud-based components, LocalStack \cite{Localstack} was used, enabling us to mimic AWS services locally.

\subsubsection{Backend Runtime}
This evaluation focuses on the runtime overhead of our backend server, including all processes from the receipt of client data in the queue to the completion of the KS calculation. The client stores $d=20,000$ samples, and only then sends them to the cloud. The reference dataset comprises randomly selected 100,000 samples from the real-world transactions dataset.

\begin{figure}[h]
\centering
\includegraphics[width=1\columnwidth]{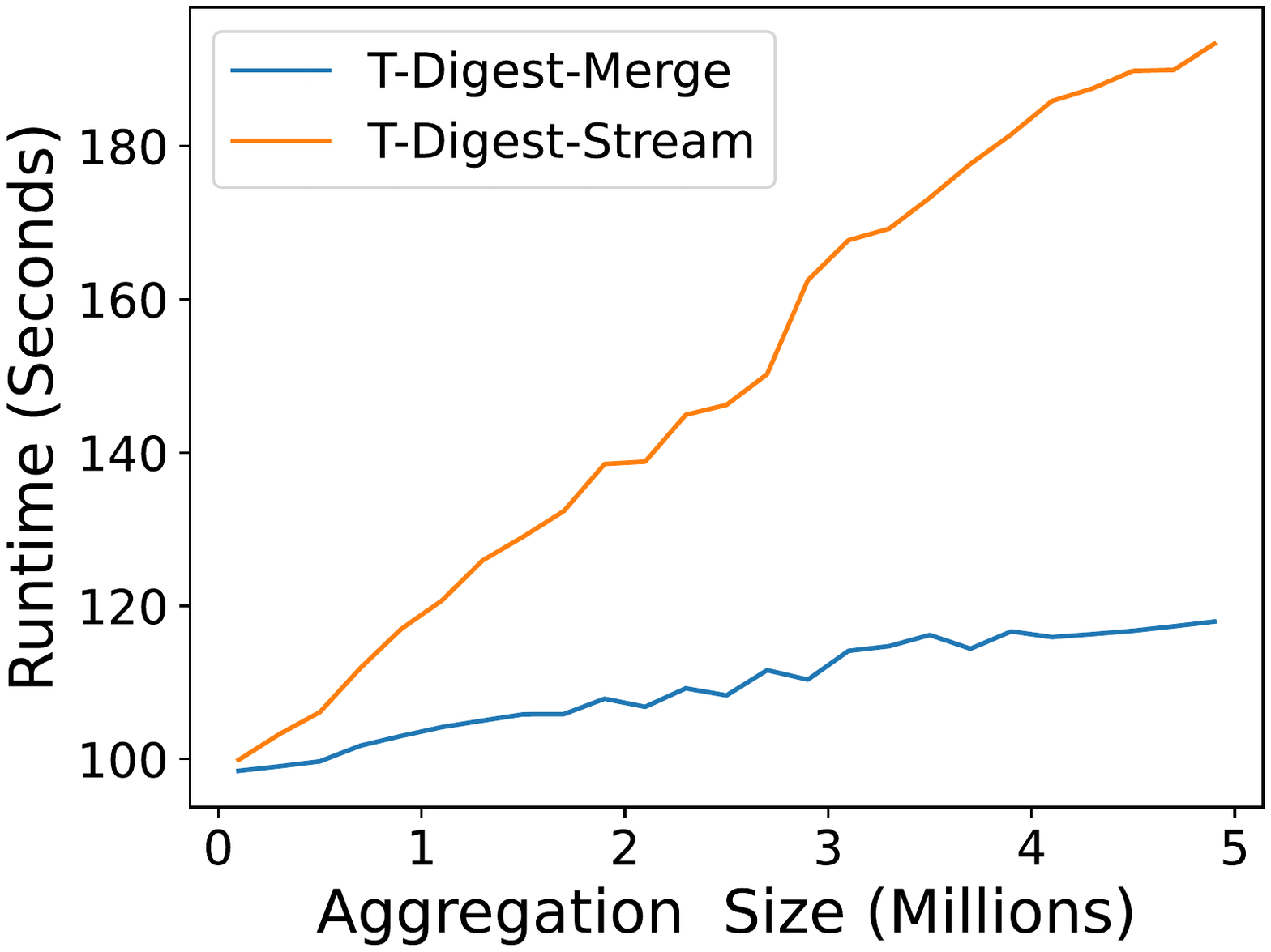}
\caption{Backend cycle runtime by server aggregation size}
\label{fig:backend_run_time}
\end{figure}

Figure \ref{fig:backend_run_time} demonstrates the backend runtime as a function of the aggregation size.
The backend's runtime scales linearly with the increase in aggregation size. This linear trend is indicative of the system's ability to scale efficiently under varying loads. Additionally, a notable improvement is observed when compared to T-Digest-Stream, which reaches a ratio of $1.5$ at aggregation size of 5 million between the T-Digest-Stream and T-Digest-Merge.

\subsubsection{Client Runtime}
Given the deployment of the client on edge devices, which have limited computational resources, our framework must operate with minimal resource consumption. This experiment compares T-Digest-Merge runtime on the client side against T-Digest-Stream, considering different client batch sizes. The client runtime includes data structure management, updating with new samples, and the processes of serialization of the data structure and transmission over the network.

\begin{figure}[h]
\centering
\includegraphics[width=1\columnwidth]{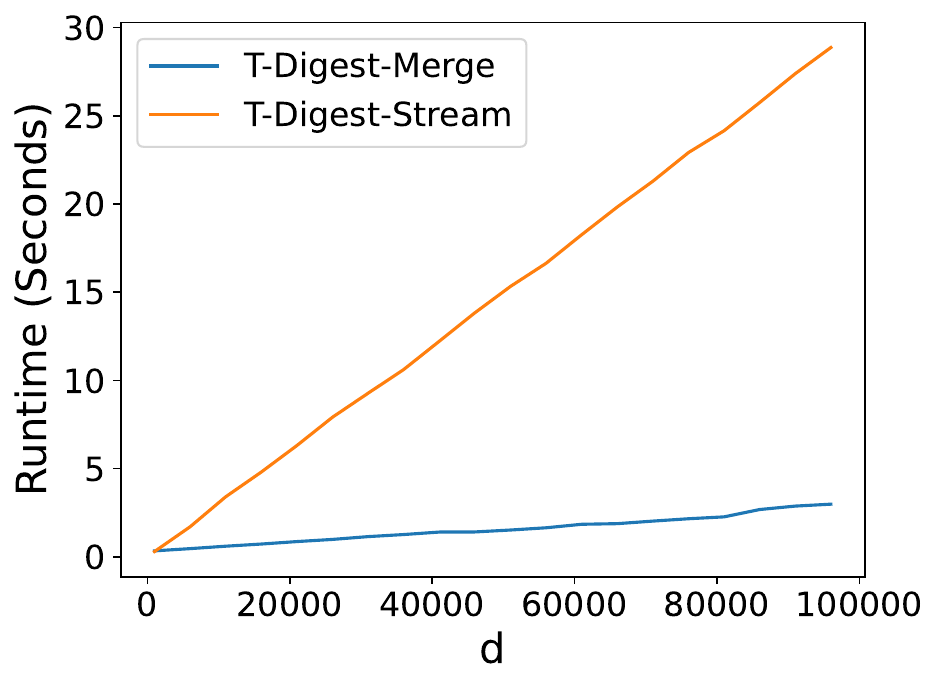}
\caption{Client runtime as a function of $d$}
\label{fig:client_run_time}
\end{figure}

Figure~\ref{fig:client_run_time} depicts the client runtime as a function of $d$, the number of samples stored at the client before sending them to the cloud. T-Digest-Merge demonstrates superior client runtime performance compared to T-Digest-Stream. Notably, even though our method involves managing the $t$-digest structure on the client, it remains more efficient overall. This efficiency is attributed to the reduced network overhead from sending less data, and to the high efficiency of the $t$-digest operations.

\section{Discussion and Future Work}
This research introduces a framework for monitoring data distribution shifts in edge ML deployments, specifically aimed to enable edge ML fraud detection. Our findings demonstrate the framework's accuracy and efficiency in real-world settings, highlighting its potential as a vital component for fraud detection using edge ML.
The rapid growth in IoT devices and the widespread adoption of smartphones with increasing capabilities underscore the significance of edge ML, particularly in the context of fraud detection. 

While edge ML is promising, it also brings forth significant challenges that require addressing. The proposed framework, though effective in certain aspects, falls short of providing a holistic solution to the array of challenges in this new paradigm. This shortfall is also seen in the fraud detection use case. Specifically, in fraud detection, it's not sufficient to monitor data distribution shifts, as the data faced is highly imbalanced, and legitimate transactions significantly outnumber fraudulent ones. Thus, shifts in the overall data distribution might overlook critical changes in fraudulent activity patterns. Such scenarios are undetectable in the proposed framework, as the data distribution shift will be negligible.

To tackle these challenges effectively, future research must focus on developing a holistic solution. One that reflects the diverse strategies used in cloud monitoring systems. A robust solution for edge ML needs to incorporate a variety of methods, creating a framework that offers operators a comprehensive overview and addresses the various obstacles. Our work is envisioned to be an integral component of these future solutions, contributing to the development of edge ML operations.

An interesting avenue for future research is extending our framework to offer privacy-preserving guarantees. Given the critical importance of privacy in certain edge ML use cases, developing methods to accurately measure and monitor collective data distributions while maintaining privacy is essential. Another interesting improvement of our system is incorporating a sliding window approach to continuously calculate the aggregated distribution. This approach would reduce the latency of detecting a distribution shift.

\section{Conclusion}
In this study, we developed a novel framework aimed at monitoring data distribution shifts in edge ML applications, with a specific focus on fraud detection. In particular, we proposed a novel implementation of the KS test over a distributed network of edge devices.
Our extensive evaluations underscore the framework's accuracy and efficiency, confirming its effectiveness for practical, real-world deployment. This research marks a step towards building a comprehensive edge ML monitoring solution.

\bibliography{aaai24}

\begin{thebibliography}{34}
\providecommand{\natexlab}[1]{#1}

\bibitem[{{AWS}(2023)}]{AWS:DRIFT_DETECTION}
{AWS}. 2023.
\newblock Detecting data drift using Amazon SageMaker.
\newblock [Online; accessed 24-November-2023].

\bibitem[{AWS(2023)}]{SQS}
AWS. 2023.
\newblock SQS.
\newblock \url{https://aws.amazon.com/sqs/}.
\newblock Accessed: Nov 24, 2023.

\bibitem[{Bosamia and Patel(2019)}]{Mobile-Wallet-Threats}
Bosamia, M.; and Patel, D. 2019.
\newblock Wallet payments recent potential threats and vulnerabilities with its possible security measures.
\newblock \emph{Int. J. Comput. Sci. Eng}, 7(1): 810--817.

\bibitem[{Chen et~al.(2019)Chen, Zhang, Liu, Yang, Meng, and Wang}]{Fraud-Detection-Based-on-Individual-Behavior}
Chen, L.; Zhang, Z.; Liu, Q.; Yang, L.; Meng, Y.; and Wang, P. 2019.
\newblock A Method for Online Transaction Fraud Detection Based on Individual Behavior.
\newblock In \emph{Proceedings of the ACM Turing Celebration Conference - China}, ACM TURC '19. New York, NY, USA: Association for Computing Machinery.
\newblock ISBN 9781450371582.

\bibitem[{Chen et~al.(2005)Chen, Luo, Liang, and Lee}]{Personalized-Detecting-Fraud}
Chen, R.-C.; Luo, S.-T.; Liang, X.; and Lee, V. 2005.
\newblock Personalized Approach Based on SVM and ANN for Detecting Credit Card Fraud.
\newblock In \emph{2005 International Conference on Neural Networks and Brain}, volume~2, 810--815.

\bibitem[{ComplyAdvantage(2023)}]{complyadvantage}
ComplyAdvantage. 2023.
\newblock complyadvantage.
\newblock \url{https://complyadvantage.com/}.
\newblock Accessed: Nov 24, 2023.

\bibitem[{Dal~Pozzolo et~al.(2015)Dal~Pozzolo, Boracchi, Caelen, Alippi, and Bontempi}]{Credit-card-fraud-detection-and-concept-drift-adaptation-with-delayed-supervised-information}
Dal~Pozzolo, A.; Boracchi, G.; Caelen, O.; Alippi, C.; and Bontempi, G. 2015.
\newblock Credit card fraud detection and concept-drift adaptation with delayed supervised information.
\newblock In \emph{2015 International Joint Conference on Neural Networks (IJCNN)}, 1--8.

\bibitem[{Dal~Pozzolo et~al.(2018)Dal~Pozzolo, Boracchi, Caelen, Alippi, and Bontempi}]{Realistic-Modeling-and-a-Novel-Learning-Strategy}
Dal~Pozzolo, A.; Boracchi, G.; Caelen, O.; Alippi, C.; and Bontempi, G. 2018.
\newblock Credit Card Fraud Detection: A Realistic Modeling and a Novel Learning Strategy.
\newblock \emph{IEEE Transactions on Neural Networks and Learning Systems}, 29(8): 3784--3797.

\bibitem[{Davidson-Pilon(2023)}]{t-digest-python}
Davidson-Pilon, C. 2023.
\newblock T-Digest data structure in Python.
\newblock \url{https://github.com/CamDavidsonPilon/tdigest}.
\newblock Accessed: 2023-11-11.

\bibitem[{Dos~Reis et~al.(2016)Dos~Reis, Flach, Matwin, and Batista}]{KS2016fast}
Dos~Reis, D.~M.; Flach, P.; Matwin, S.; and Batista, G. 2016.
\newblock Fast unsupervised online drift detection using incremental kolmogorov-smirnov test.
\newblock In \emph{Proceedings of the 22nd ACM SIGKDD international conference on knowledge discovery and data mining}, 1545--1554.

\bibitem[{dos Reis et~al.(2016)dos Reis, Flach, Matwin, and Batista}]{Online-Drift-Detection-Using-Incremental-KS}
dos Reis, D.~M.; Flach, P.; Matwin, S.; and Batista, G. 2016.
\newblock Fast Unsupervised Online Drift Detection Using Incremental Kolmogorov-Smirnov Test.
\newblock In \emph{Proceedings of the 22nd ACM SIGKDD International Conference on Knowledge Discovery and Data Mining}, KDD '16, 1545–1554. New York, NY, USA: Association for Computing Machinery.
\newblock ISBN 9781450342322.

\bibitem[{Dunning and Ertl(2019)}]{t-digest}
Dunning, T.; and Ertl, O. 2019.
\newblock Computing extremely accurate quantiles using t-digests.
\newblock \emph{arXiv preprint arXiv:1902.04023}.

\bibitem[{Featurespace(2023)}]{featurespace}
Featurespace. 2023.
\newblock featurespace.
\newblock \url{https://www.featurespace.com/}.
\newblock Accessed: Nov 24, 2023.

\bibitem[{{GCP}(2023)}]{GCP:DRIFT_DETECTION}
{GCP}. 2023.
\newblock Monitor feature skew and drift.
\newblock [Online; accessed 24-November-2023].

\bibitem[{Hao et~al.(2023)Hao, Wang, Hong, Li, Karayanni, Mao, Yang, and Cidon}]{nazar}
Hao, W.; Wang, Z.; Hong, L.; Li, L.; Karayanni, N.; Mao, C.; Yang, J.; and Cidon, A. 2023.
\newblock Monitoring and Adapting ML Models on Mobile Devices.
\newblock \emph{arXiv preprint arXiv:2305.07772}.

\bibitem[{Hendrycks and Gimpel(2017)}]{Confidence-for-drift-detection}
Hendrycks, D.; and Gimpel, K. 2017.
\newblock A Baseline for Detecting Misclassified and Out-of-Distribution Examples in Neural Networks.
\newblock In \emph{International Conference on Learning Representations}.

\bibitem[{Lall(2015)}]{streaming-KS}
Lall, A. 2015.
\newblock Data streaming algorithms for the Kolmogorov-Smirnov test.
\newblock In \emph{2015 IEEE International Conference on Big Data (Big Data)}, 95--104.

\bibitem[{Loxton et~al.(2020)Loxton, Truskett, Scarf, Sindone, Baldry, and Zhao}]{Consumer-Behaviour-during-Crises}
Loxton, M.; Truskett, R.; Scarf, B.; Sindone, L.; Baldry, G.; and Zhao, Y. 2020.
\newblock Consumer Behaviour during Crises: Preliminary Research on How Coronavirus Has Manifested Consumer Panic Buying, Herd Mentality, Changing Discretionary Spending and the Role of the Media in Influencing Behaviour.
\newblock \emph{Journal of Risk and Financial Management}, 13(8).

\bibitem[{Massey(1951)}]{KS-Original-Paper}
Massey, F.~J. 1951.
\newblock The Kolmogorov-Smirnov Test for Goodness of Fit.
\newblock \emph{Journal of the American Statistical Association}, 46(253): 68--78.

\bibitem[{Mishra(2023)}]{Localstack}
Mishra, H. 2023.
\newblock LocalStack's Docker Extension.
\newblock \url{https://blog.localstack.cloud/2023-01-13-introducing-localstack-extension-for-docker-desktop/}.
\newblock Accessed: Nov 24, 2023.

\bibitem[{NannyML(2023)()}]{nannyml}
NannyML(2023). 2023.
\newblock {N}anny{ML} (release 0.9.1).
\newblock \url{https://github.com/NannyML/nannyml}.
\newblock Accessed: Nov 24, 2023.

\bibitem[{Nguyen(2018)}]{big_data_ks}
Nguyen, H.~D. 2018.
\newblock A Two-Sample Kolmogorov-Smirnov-Like Test for Big Data.
\newblock In Boo, Y.~L.; Stirling, D.; Chi, L.; Liu, L.; Ong, K.-L.; and Williams, G., eds., \emph{Data Mining}, 89--106. Singapore: Springer Singapore.
\newblock ISBN 978-981-13-0292-3.

\bibitem[{Omar, Fred, and Swaib(2018)}]{state-of-the-art-review-of-machine-learning-techniques-for-fraud-detection}
Omar, S.~J.; Fred, K.; and Swaib, K.~K. 2018.
\newblock A State-of-the-Art Review of Machine Learning Techniques for Fraud Detection Research.
\newblock In \emph{Proceedings of the 2018 International Conference on Software Engineering in Africa}, SEiA '18, 11–19. New York, NY, USA: Association for Computing Machinery.
\newblock ISBN 9781450357197.

\bibitem[{Pozzolo et~al.(2015)Pozzolo, Caelen, Johnson, and Bontempi}]{real-world-ds}
Pozzolo, A.~D.; Caelen, O.; Johnson, R.~A.; and Bontempi, G. 2015.
\newblock Calibrating Probability with Undersampling for Unbalanced Classification.
\newblock In \emph{2015 IEEE Symposium Series on Computational Intelligence}, 159--166.

\bibitem[{RB and KR(2021)}]{Credit-card-fraud-detection-using-artificial-neural-network}
RB, A.; and KR, S.~K. 2021.
\newblock Credit card fraud detection using artificial neural network.
\newblock \emph{Global Transitions Proceedings}, 2(1): 35--41.
\newblock 1st International Conference on Advances in Information, Computing and Trends in Data Engineering (AICDE - 2020).

\bibitem[{{Redis}(2023)}]{Redis}
{Redis}. 2023.
\newblock Redis.io.
\newblock [Online; accessed 24-November-2023].

\bibitem[{Roy(2021)}]{generatedfraud186}
Roy, R. 2021.
\newblock Online Payments Fraud Detection Dataset.
\newblock \url{https://www.kaggle.com/datasets/rupakroy/online-payments-fraud-detection-dataset/data}.

\bibitem[{Sharma, Sharma, and Lal(2019)}]{Anomaly-Detection-Techniques:ON-DEVICE-ML-FOR-FRAUD-DETECTION}
Sharma, B.; Sharma, L.; and Lal, C. 2019.
\newblock Anomaly Detection Techniques using Deep Learning in IoT: A Survey.
\newblock In \emph{2019 International Conference on Computational Intelligence and Knowledge Economy (ICCIKE)}, 146--149.

\bibitem[{Shenoy(2020)}]{generatedfraud212}
Shenoy, K. 2020.
\newblock Credit Card Transactions Fraud Detection Dataset.
\newblock \url{https://www.kaggle.com/datasets/kartik2112/fraud-detection/}.

\bibitem[{Wang, Hahn, and Sutrave(2016)}]{Mobile_payment_security_threats_and_challenges}
Wang, Y.; Hahn, C.; and Sutrave, K. 2016.
\newblock Mobile payment security, threats, and challenges.
\newblock In \emph{2016 Second International Conference on Mobile and Secure Services (MobiSecServ)}, 1--5.

\bibitem[{Wang et~al.(2022)Wang, Wang, Zhang, Zhan, Guo, Zheng, and Wang}]{A-survey-on-deploying-mobile-deep-learning-applications}
Wang, Y.; Wang, J.; Zhang, W.; Zhan, Y.; Guo, S.; Zheng, Q.; and Wang, X. 2022.
\newblock A survey on deploying mobile deep learning applications: A systemic and technical perspective.
\newblock \emph{Digital Communications and Networks}, 8(1): 1--17.

\bibitem[{Widmer and Kubat(1994)}]{LearninginthePresence-of-Drift}
Widmer, G.; and Kubat, M. 1994.
\newblock Learning in the Presence of Concept Drift and Hidden Contexts.
\newblock \emph{Machine Learning}, 23.

\bibitem[{{Wikipedia}(2023)}]{enwiki:ks}
{Wikipedia}. 2023.
\newblock Kolmogorov–Smirnov test --- {Wikipedia}{,} The Free Encyclopedia.
\newblock [Online; accessed 24-November-2023].

\bibitem[{Yang et~al.(2019)Yang, Zhang, Ye, Li, and Xu}]{FFD}
Yang, W.; Zhang, Y.; Ye, K.; Li, L.; and Xu, C.-Z. 2019.
\newblock FFD: A Federated Learning Based Method for Credit Card Fraud Detection.
\newblock In Chen, K.; Seshadri, S.; and Zhang, L.-J., eds., \emph{Big Data -- BigData 2019}, 18--32. Cham: Springer International Publishing.
\newblock ISBN 978-3-030-23551-2.

\end{thebibliography}

\end{document}